\documentclass[preprint,aps]{revtex4}
\usepackage{graphicx}%

\begin{document}
\title{Possible Connection between Probability, \\
Spacetime Geometry and Quantum Mechanics}
\author{Enrique Canessa\footnote{e-mail: canessae@ictp.it}}
\affiliation{The Abdus Salam International Centre for Theoretical Physics,
Trieste, Italy
\vspace{3cm}
}

\begin{abstract}
Following our discussion [Physica A {\bf 375} (2007) 123] to associate 
an analogous probabilistic description with spacetime geometry in 
the Schwarzschild metric from the macro- to the micro-domain, we 
argue that there is a possible connection among normalized 
probabilities ${\cal P}$, spacetime geometry
(in the form of Schwarzschild radii $r_{s}$) and quantum mechanics
(in the form of complex wave functions $\psi$), namely
$\sqrt{{\cal P}^{(n)}_{\theta,\phi,t}}
     \approx R^{(n)}_{s} / r_{s} =
|\psi^{(n)}_{n}(X^{(n)})|^{2} / |\psi_{n}(x)|^{2}$.
We show how this association along different $(n)$-nested surfaces
--representing curve space due to an inhomogeneous density of matter--
preserves the postulates of quantum mechanics at different
geometrical scales.
\end{abstract}
\maketitle

\section{Overview}

In a recent paper (hereafter denoted as I) \cite{Can07}, we argued 
how the universe we live in (consisting of an inhomogeneous density 
of matter in the form of starts, molecules, atoms, {\it etc}, that 
curve space due to their gravitational fields) may be seen 
fullfilled with nested surfaces.  In I we considered 
Schwarzschild's isotropic metric --which is the base for tests of 
general relativity and of the existence of black holes-- in the vicinity
of multiple massive objects.  Within this picture, we demonstrated that
the probabilistic description of stochastic processes of a general 
{\it birth-and-death} model can be associated to the Schwarzschild 
metric at different geometrical scales. 

Inspired by this idealization (which in I it was considered a toy
model since there were no output variables), our goal in this work 
is to argue that there is a possible connection between
an analogous probability, 
spacetime curvature and quantum mechanics.  
Within this association, the postulates of quantum mechanics are
preserved on scales ranging from the size of the observable universe to 
micro-world distances greater than the black hole horizon.  We use same
notation as in I and write down the algebra in details.

\section{probability and spacetime geometry}

Let us start with the standard isotropic form of Schwarzschild's 
metric \cite{Fos94}
\begin{equation}\label{eq:schwarzschild}
ds^{2} = \left(1-{2m \over r}\right)\;c^{2}dt^{2} -
     \left(1-{2m \over r}\right)^{-1} dr^{2}-r^{2}d\Omega^{2}\;\;\; ,
\end{equation}
where $m \equiv GM/c^{2}$, $M$ is the mass of the body producing
the field, $G$ is the Gravitational constant, $c$ is the speed of
light and $d\Omega^{2} = d\theta^{2}+sin \theta \; d\phi^{2}$ is the 
element of solid angle.  
Infinitesimal radial distances in this metric, at 
fixed polar angles $\theta$ and $\phi$ and time $t$, then satisfy 
\begin{equation}\label{eq:dr}
dR \equiv \left(1-{2m \over r}\right)^{-1/2} dr \;\;\; .
\end{equation}
In I, we extended this equation to a system of nested curved surfaces 
$S_{m}$ forming a spiral of $n$-interconnected curved surfaces 
along distance scales from the macro- to the micro-world. 
In other words we assumed the space to be curved 
at all geometric scales $dR^{(n)}$ due to the presence of 
scattered matter $M^{(i)} \ne 0 \;(i=0,1,\cdots,n$) with 
$M^{(0)}\equiv m$.  We then established the master relation
\begin{equation}\label{eq:master}
\left( { dr \over dR^{(n)} } \right)^{2} \equiv 
           {\cal P}^{(n)}_{\theta,\phi,t}(r) < 1 \;\;\; .
\end{equation}
This relates space 
geometry to an iterative ($n$-process) probability-like function.
We proved rigorously in I that the function ${\cal P}^{(n)}$ 
can be considered as an analogous probability measure since 
it implies well-defined stationary states to exist and
its sum satisfies the normalization condition
$\sum_{i=0}^{n}{\cal P}^{(i)}=\sum_{i=0}^{n}\left(dr/dR^{(i)}\right)^{2}\equiv 1$
over $n$ different states.

The analogous probabilities ${\cal P}$, at fixed polar angles 
and time, are obtained recursively from the {\it birth-and-death}-like 
general recursive relation 
\begin{equation}\label{eq:bd}
{\cal P}^{(n+1)}(r) = \left( { \lambda^{(n)} \over \mu^{(n+1)} }\right) \; {\cal P}^{(n)}(r) \rightleftharpoons \left( 1 - { R^{(n+1)}_{s} \over R^{(n)} } \right) {\cal P}^{(n)}(r) \;\;\; ,
\end{equation}
such that
\begin{eqnarray}\label{eq:lambdamu}
{\cal P}^{(n)}(r) & \rightleftharpoons &
{ \lambda^{(n-1)}\cdots \lambda^{(1)} \lambda^{(0)} \over \mu^{(n)}\cdots
\mu^{(2)}\mu^{(1)} }
\; {\cal P}^{(0)}(r) \;\;\; , \nonumber \\
{\cal P}^{(0)}(r) & \rightleftharpoons &  1 - { r_{s} \over r } \;\;\; , \nonumber \\
\lambda^{(n-1)}  & \rightleftharpoons & ( R^{(n-1)} - R^{(n)}_{s} )/r_{s}  \;\;\; \nonumber \\
\mu^{(n)}  & \rightleftharpoons &  R^{(n-1)}/r_{s} \;\;\; .
\end{eqnarray}
In the above $R^{(n)}_{s} = 2GM^{(n)}/c^{2}$ are Schwarzschild radii 
--at which the metric of eq.(\ref{eq:dr}) becomes singular for positive
$n \ge 1$, $R^{(0)}_{s} \equiv r_{s} = 2m$, $R^{(0)} \equiv R$, and
$(\lambda,\mu)$ are {\it birth-and-death} coefficients, respectively.

Using forward and backward two-point 
approximations for the derivative of the scattered 
matter $M^{(n+1)}-M^{(n)} \approx dM^{(n)}/dn$ and
$(\mu^{(n+1)} - \mu^{(n)}) - (\lambda^{(n)} -\lambda^{(n-1)}) \approx
(d\mu^{(n)}/dn) - (d\lambda^{(n)}/dn)$,
in I we also suggested that 
\begin{equation}\label{eq:mass}
\frac{1}{M}\left(\frac{dM^{(n)}}{dn}\right) \approx
\frac{1}{r_{s}}\left(\frac{dR_{s}^{(n)}}{dn}\right) 
 \rightleftharpoons  
\frac{d\mu^{(n)}}{dn} - \frac{d\lambda^{(n)}}{dn}
\approx \frac{1}{{\cal E}}\left(\frac{d{\cal E}^{(n)}}{dn}\right)
  \;\;\; .
\end{equation}
These relations imply that variations of an analogous system energy
(and mass) coefficient can be related to: {\it i}) the 
difference between some 'annihilated' analogous processes of 
leaving state $n$ and those being 'generated' in the state $n$, 
or {\it ii}) some small (stochastic) fluctuations in the 
Schwarzschild radii.

\section{Possible Connection}

Let us extend the main ideas behind eq.(\ref{eq:master}) --{\it i.e.}, 
the concept of analogous probability as being consistent with the 
full filling of
the curved space with nested surfaces-- to a general (non necessarily 
stochastic) 1D process.  We keep $r$ and $R$ as independent variables 
at different geometrical scales correlated by the Schwarzschild metric.
We also consider $P^{(0)}(r) = constant$, which is the case in 
eq.(\ref{eq:lambdamu}) for distances greater than the Schwarzschild 
radii $r_{s}$.

The coordinate distance (length) is here approximated to lie radially 
in the field of each $n$-spherical object of mass $M^{(n)}$ 
(see, {\it e.g.}, \cite{Fos94} ), namely
$dr \rightarrow \Delta r = x - x_{0}$. At each nested surface
we approximate
$dR^{(n)} \rightarrow \Delta R^{(n)} = X^{(n)} - X^{(n)}_{0}$.
Our master equation therefore simplifies to
\begin{equation}\label{eq:master2}
 { dr \over dR^{(n)} } \rightarrow { \Delta r \over \Delta R^{(n)} } 
 = { x - x_{0} \over X^{(n)} - X^{(n)}_{0}} \equiv \sqrt{ {\cal P}^{(n)} }
   < 1 \;\;\; .
\end{equation}

For $\Delta$'s small enough, and choosing the origin at 
$x_{0}=X^{(n)}_{0}\equiv 0$, we can estimate the ratio of
(micro to macro) potential fields as
\begin{equation}\label{eq:potential}
 { dr \over dR^{(n)} }  \rightarrow
  { \Delta r \over \Delta R^{(n)} }  \approx 
 \left( { x \over - GMM^{(n)} }\right) \cdot \left( { - GMM^{(n)} \over X^{(n)} }\right)
   \equiv \frac{U^{(n)}}{U}  = \sqrt{{\cal P}^{(n)}}   \;\;\; , 
\end{equation}
with $U \ne 0$.
The system masses are independent of $x$ as deduced from eq.(\ref{eq:mass}).  
That is, 
\begin{equation}\label{eq:mass1}
\frac{M^{(n)}}{M} \approx { R^{(n)}_{s} \over r_{s} } \rightleftharpoons  
  \mu^{(n)} - \lambda^{(n)} \approx 
        \frac{{\cal E}^{(n)}}{{\cal E}} \;\;\; . 
\end{equation}
Any constant factor in the integration is taken to be zero.

For further progress, it is worth to check our associations against
quantum physics measurements, at least up 
to a submolecular magnitude of the order of the Bohr radius 
$x\rightarrow a_{0}=\hbar^{2}/me^{2} \sim 0.529 \AA$ 
\cite{Blin04,Coh77}.  Considering circular orbits in the Bohr atom model,  
the quantized angular-momentum is assumed to be $L\equiv n\hbar$ (with
$L^{2}=me^{2}r$), and the allowed orbital radii are given 
by $r_{n} = n^{2} a_{0}$.  The corresponding energy is 
$E_{n} = - e^{2}/2a_{0}n^{2}$ with $n=1,2, \cdots$   
These relations can be rewritten as
\begin{equation}\label{eq:bohr}
  \left({ a_{0} \over r_{n} }\right)^{2} = 
     \left({ r_{1} \over r_{n} }\right)^{2} = { 1 \over n^{4} } 
    \;\;\; ; \;\;\; { E_{n} \over E_{1} } =  { 1 \over n^{2} }  \;\;\; .
\end{equation}
In conjunction with our eqs.(\ref{eq:master2}) and (\ref{eq:mass1}),
for $n>1$ we can readily identify 
\begin{equation}\label{eq:mass2}
\left({ x \over X^{(n)} }\right)^{2} = 
   {\cal P}^{(n)} \propto { 1 \over n^{4} } < 1 \;\;\; ; \;\;\; 
  \frac{M^{(n)}}{M} \approx \frac{{\cal E}^{(n)}}{{\cal E}} \rightarrow
        { 1 \over n^{2} } \propto \sqrt{ {\cal P}^{(n)} }  \;\;\; , 
\end{equation}
which leads the association $M^{(n)}/M \rightleftharpoons x/X^{(n)}$. 
This, in  turn, implies $M^{(n)}e^{2}X^{(n)} \rightleftharpoons Me^{2}x$ which
means that the angular momentum is conserved. 
An estimate of $L$ can be obtained by considering circular 
orbits of radius $X^{(n)}$ around $x \rightarrow a_{0}$, {\it i.e.}, 
$L^{2}=Me^{2}X^{(n)}= (\hbar^{2}/a_{0})(x/\sqrt{{\cal P}^{(n)}})$.
In conjunction with eq.(\ref{eq:mass2}), the angular-momentum quantization 
becames $L=\hbar/({\cal P}^{(n)})^{1/4} \rightarrow n\hbar$.
This is just an illustrative example since
Schwarzschild's metric is valid for non-rotating bodies only.
The results of the Bohr empirical model given in eq.(\ref{eq:mass2}) are 
our motivation though to depict the general Hamiltonian treatment of 
quantum mechanics.

Let us derive the kinetics energy operator 
$\hat{K}\equiv -(\hbar^{2}/2M)(d^{2}/dx^{2})$
in 1D, by applying again eq.(\ref{eq:master}) and the second derivative 
of $x$ discussed in the Appendix.  Using also eq.(\ref{eq:mass2}) for the 
scattered mass, it follows then that
\begin{equation}\label{eq:kinetics}
\hat{K} \equiv
{ -\hbar^{2} \over 2M } \left( { d^{2} \over dx^{2} } \right)  = 
{ -\hbar^{2} \sqrt{ {\cal P}^{(n)} } \over 2M^{(n)} } \cdot
  { 1 \over {\cal P}^{(n)} } \left( { d^{2} \over dX^{(n)2} } \right) =
{ -\hbar^{2} \over \sqrt{ {\cal P}^{(n)} } 2M^{(n)} }
 \left( { d^{2} \over dX^{(n)2} } \right) 
  \equiv { \hat{K}^{(n)} \over \sqrt{ {\cal P}^{(n)} } } \;\;\; .
\end{equation}
This association, together with eq.(\ref{eq:potential}), leads to 
\begin{equation}\label{eq:totenergy}
  \hat{{\cal E}}^{(n)} \equiv \hat{K}^{(n)} + U^{(n)} \rightarrow 
     \sqrt{{\cal P}^{(n)}} \; (\hat{K} + U) = 
             \sqrt{{\cal P}^{(n)}} \; \hat{{\cal E}} \;\;\; .
\end{equation}
A result that extends eq.(\ref{eq:mass2}) for the total system 
energy ${\cal E}$ to an energy operator at different iterations.
Therefore from the definition of the time-independent 
quantum mechanics Hamiltonian
$\hat{K}\psi_{n}+U\psi_{n}\equiv\hat{H}\psi_{n}=\hat{{\cal E}}_{n}\psi_{n}$,
with $\psi (x)$ a quantized system wave function and $\hat{{\cal E}}_{n}$
representing energy states, we see that
\begin{equation}\label{eq:totenergy1}
 \hat{{\cal E}}_{n}^{(n)}\psi^{(n)}_{n} = \hat{H}^{(n)}\psi^{(n)}_{n} 
     \rightarrow \hat{H}\psi_{n} = \hat{{\cal E}}_{n}\psi_{n} \;\;\; ,
\end{equation}
provided the wave function in each nested surface satisfies, 
$\psi^{(n)}_{n}(X^{(n)}) \rightleftharpoons \Lambda^{(n)} \psi_{n}(x)$.
(In the right hand side of eq.(\ref{eq:totenergy}) the factor
$\sqrt{{\cal P}^{(n)}}$ cancels out and $\Lambda$ is
a non zero function to be defined next).  $\psi$ is allowed to be complex.

Using the probability density postulate of quantum mechanics 
and the relation between $dx$ and $dX^{(n)}$ of eq.(\ref{eq:master2}), 
we immediately get
\begin{equation}\label{eq:probdensity}
d\Pi_{n} = |\psi_{n}(x)|^{2} dx \rightleftharpoons 
    \left|{ \psi^{(n)}_{n}(X^{(n)}) \over \Lambda^{(n)} }\right|^{2}
           \sqrt{{\cal P}^{(n)}} dX^{(n)} =
    { \sqrt{{\cal P}^{(n)}} \over \Lambda^{(n)2} }
   \; d\Pi^{(n)}_{n} \;\;\; .
\end{equation}
Therefore in order to preserve both the general Hamiltonian
form of eq.(\ref{eq:totenergy1}) and the probability density 
$\Pi_{n} = \Pi^{(n)}_{n}$ --at the n-$essima$ nested surface and at
the same quantum state $n$,  we must to have
\begin{equation}\label{eq:psin}
\Lambda^{(n)} = ({ \cal P}^{(n)})^{1/4} 
  \rightleftharpoons \psi^{(n)}_{n}(X^{(n)})/\psi_{n}(x) \;\;\; .  
\end{equation}
This, in turn, leads to the correct mathematical requirement of normalization
for the wave functions
\begin{equation}\label{eq:normalization}
\int |\psi_{n}(x)|^{2} \; dx  \rightleftharpoons \int 
\left|{ \psi^{(n)}_{n}(X^{(n)}) \over ({ \cal P}^{(n)})^{1/4} }\right|^{2}
   \sqrt{{\cal P}^{(n)}} dX^{(n)} = 
   \int |\psi^{(n)}_{n}(X^{(n)})|^{2} \; dX^{(n)} \equiv 1 \;\;\; . 
\end{equation}

For a system in a state described by the normalized wave function 
above, the expectation value of an observable corresponding to $A$ 
is also conserved independent of the chosen nested surface.  In fact,
\begin{eqnarray}\label{eq:expectation}
<A> \equiv \int \psi_{n}^{*}(x)\hat{A}\psi_{n}(x) \; dx & \rightleftharpoons  &
      \int \left({ \psi^{(n)*}_{n}(X^{(n)}) \over ({ \cal P}^{(n)})^{1/4} }\right) \hat{A} 
            \left({ \psi^{(n)}_{n}(X^{(n)}) \over ({ \cal P}^{(n)})^{1/4} }\right) \sqrt{{\cal P}^{(n)}} \; dX^{(n)}  \nonumber \\ 
    & & 
   = \int \psi^{(n)*}_{n}(X^{(n)})\hat{A}\psi^{(n)}_{n}(X^{(n)}) \; dX^{(n)} \;\;\; . 
\end{eqnarray}

Finally, it can be seen  that the property of orthogonality (a general 
result for quantum mechanical eigenfunctions \cite{Blin04,Coh77}) is 
also preserved
\begin{eqnarray}\label{eq:orthogonality}
\int \psi_{m}(x)\psi_{n}(x)\; dx & = & \delta_{m,n} \rightleftharpoons 
\int \left({ \psi^{(n)}_{m}(X^{(n)}) \over ({ \cal P}^{(n)})^{1/4} }\right) \left({ \psi^{(n)}_{n}(X^{(n)}) \over ({ \cal P}^{(n)})^{1/4} }\right)
   \sqrt{{\cal P}^{(n)}} \; dX^{(n)} \nonumber \\
    & & \hspace{1cm} 
  = \int \psi^{(n)}_{m}(X^{(n)})\psi^{(n)}_{n}(X^{(n)}) \; dX^{(n)}
     \;\;\; .
\end{eqnarray}

\section{Remarks}

In order to verify that the present connection between an
analogous probability, spacetime 
geometry and quantum mechanics makes sense,
let us consider the results of the Schr\"{o}dinger equation
for a particle of mass $M$ in a box of length $L$.  This 
simplest nontrivial model, having potential energy 
$U = U^{(n)} \equiv 0$, illustrates many of the fundamental 
concepts of quantum mechanics \cite{Blin04,Coh77}.  
The model predictions are:
energy levels $E_{n} = (h^{2}/8ML^{2})n^{2}$ 
and normalized eigenfunctions  
$\psi_{n}(x) = (2/L)^{1/2} sin (n\pi x/L)$, with $n=1,2, \cdots$

By applying our associations such that
$L/L^{(n)} \rightarrow x/X^{(n)} = \sqrt{{\cal P}^{(n)}} \rightleftharpoons M^{(n)}/M$, it 
follows that for the particle in a box
\begin{equation}\label{eq:psibox}
\psi_{n}(x) = \left( { 2 \over \sqrt{{\cal P}^{(n)}} L^{(n)}} 
                      \right)^{1/2} 
   sin \left( { n\pi \sqrt{{\cal P}^{(n)}} X^{(n)} \over \sqrt{{\cal P}^{(n)}} L^{(n)} } \right) 
   = { \psi_{n}(X^{(n)}) \over ({\cal P}^{(n)})^{1/4} } \;\;\; ,
\end{equation}
and
\begin{equation}\label{eq:ebox}
E_{n} = \left( { \sqrt{{\cal P}^{(n)}} h^{2} \over 8 M^{(n)} 
          (\sqrt{{\cal P}^{(n)}} L^{(n)})^{2} } \right) n^{2} =
     { E^{(n)}_{n} \over \sqrt{{\cal P}^{(n)}} } \;\;\; .
\end{equation}
Both of these relations are compatible with eqs.(\ref{eq:psin}) and
(\ref{eq:totenergy}), respectively.  These results emphasize our
suggestion to perceive spacetime in terms of surfaces interconnected 
due to the presence of dispersed mass even at atomic levels.  From 
eq.(\ref{eq:mass2}), we have in our notation that 
$M^{(0)} > M^{(1)} > \cdots > M^{(n)}$ (hence, $x < X^{(n)}$) and
$dr < dR < dR^{(1)} < \cdots < dR^{(n)}$.
This simply means that the system of nested curved 
surfaces form a spiral ({\it c.f.}, Fig. 1 in I).

The normalized analogous probabilities ${\cal P}^{(n)}$ in 
eq.(\ref{eq:master}) are valid for fixed time and polar angles.  
According to eqs.(\ref{eq:mass1}) and (\ref{eq:mass2}) 
--or eq.(\ref{eq:totenergy}), we have that 
${\cal P}^{(n)} = (\mu^{(n)} - \lambda^{(n)})^{2}$ when 
considering general stochastic {\it birth-and-death} processes.
Furthermore, we have considered $P^{(0)}$ to be constant in order to 
evaluate derivatives --{\it e.g.}, those in the Appendix. This
is the case in eq.(\ref{eq:lambdamu}) for $r > r_{s}$.
Here the label $n$ relates quantum states and the upper symbol
$(n)$ accounts for the n-$essima$ interconnected surface.
The meaning of $P^{(n)}$ is different from the meaning of $\Pi^{(n)}$.  
The latter relates the probability density postulate of 
quantum mechanics in eq.(\ref{eq:probdensity}). 
Throughout different geometrical scales, we have explicitly shown that
our associations lead to preserve the Hamiltonian form 
({\it c.f.}, eq.(\ref{eq:totenergy1})) and postulates 
of quantum mechanics 
({\it c.f.}, eqs.(\ref{eq:normalization})-(\ref{eq:orthogonality}))

To summarize, we derived along different $(n)$-nested
surfaces the following novel connection
among normalized analogous probabilities, spacetime geometry 
(in the form of Schwarzschild radii) and quantum mechanics (in 
the form of complex wave functions) 
\begin{equation}\label{eq:vinculum}
  \sqrt{{\cal P}^{(n)}_{\theta,\phi,t}} 
     \approx { R^{(n)}_{s} \over r_{s} } =
{ |\psi^{(n)}_{n}(X^{(n)})|^{2} \over |\psi_{n}(x)|^{2} } \;\;\; .
\end{equation}
We believe this association could be useful
to analyze quantum mechanics processes above the event horizon, 
or quantum systems that are macroscopic both in their
spatial dimensions and in the number of particles involved 
\cite{Kon2006}, each of which causes curvature of the spacetime
around it.

At the apparent horizon of a black hole that has the Schwarzschild 
metric, or inside and around the horizon (where the role of time and 
space coordinates is interchanged), quantum fluctuations are involved
in the process of Hawking radiation \cite{Ken98}.
It is interesting to note that Hawkings predicted that a black hole
radiates thermally like a hot coal, with a temperature $T$ inversely
proportional to its mass.  For a black hole of solar mass
($M =1.99 \times 10^{30} Kg$), this implies $T \sim 10^{-6} K$ --which 
is negligible at the present age of the universe \cite{Car05}.
But for a black hole of a mountain of mass $M^{(n)} \sim 10^{12} Kg$, 
$T^{(n)} \sim 10^{12} K$ which is hot enougth to emit photons, electrons 
and positrons.  Hence we estimate the same order of magnitude
({\it i.e.}, $10^{-18}$) for
$R^{(n)}_{s}/R_{s}=M^{(n)}/M\sim 10^{12}/10^{30} \rightleftharpoons (1/T^{(n)})/(1/T) \sim 10^{-6}/10^{12}$.
This could be another compelling reason for interest in the 
present work.

\section*{Appendix}

To obtain the kinetics energy operator $\hat{K}$ given 
in eq.(\ref{eq:kinetics}) we use eq.(\ref{eq:master}) 
and consider the second derivative
\begin{eqnarray}\label{eq:secondderiv}
{ d^{2} \over dx^{2} } & = &
    { d \over dx } \left( { d \over dx } \right) =
    { d \over \left( { dx \over dX^{(n)} } \right) dX^{(n)} } 
 \left( { d \over \left( { dx \over dX^{(n)} } \right) dX^{(n)} } 
  \right) \nonumber \\  
   & & \hspace{1cm} \approx 
{ 1 \over \sqrt{ {\cal P}^{(n)} } } { d \over dX^{(n)} } 
 \left( { 1 \over \sqrt{ {\cal P}^{(n)} } } { d \over dX^{(n)} } \right) =
{ 1 \over {\cal P}^{(n)} } \left( { d^{2} \over dX^{(n)2} } \right)
\;\;\; .
\end{eqnarray}
The last term is obtained from the fact that we are deriving
with respect to the independent variable $X^{(n)}$ and that 
the analogous probability ${\cal P}$-function depends on $X^{(n-1)}$ 
(as can be deduced from eq.(\ref{eq:lambdamu})).  A similar
relation follows by applying eq.(\ref{eq:master}) directly 
to $d^{2}/dx^{2} = d^{2}/({\cal P}^{(n)}dX^{(n)2})$.

\end{document}